\documentclass{aastex}          
\usepackage{spr-astr-addons}    

\usepackage{natbib}
\usepackage{graphicx}
\usepackage{multirow}
\usepackage{txfonts}

\begin{document}
%
\title{Properties of the Black Hole Candidate XTE J1118+480 with the TCAF solution during its Jet Activity Induced 2000 Outburst}

\shorttitle{XTE J1118+480: Analysis with the TCAF Solution}
\shortauthors{Chatterjee, Debnath, Jana \& Chakrabarti}


\author{Debjit Chatterjee\altaffilmark{1}, Dipak Debnath\altaffilmark{1} and Arghajit Jana\altaffilmark{1}} 
\affil{Indian Centre for Space Physics, 43 Chalantika, Garia St. Rd., Kolkata, 700084, India.\\ email (Dipak Debnath): dipakcsp@gmail.com}
\email{dipakcsp@gmail.com}


\and

\author{Sandip K. Chakrabarti\altaffilmark{1,2}}
\affil{S. N. Bose National Centre for Basic Sciences, Salt Lake, Kolkata, 700106, India.}


\altaffiltext{1}{Indian Centre for Space Physics, 43 Chalantika, Garia St. Rd., Kolkata, 700084, India.}
\altaffiltext{2}{S. N. Bose National Centre for Basic Sciences, Salt Lake, Kolkata, 700106, India.}

\begin{abstract}

Galactic black hole candidate (BHC) XTE~J1118+480 during its 2000 outburst has been studied in a broad energy range 
using the archival data of PCA and HEXTE payloads of {\it Rossi X-ray Timing Explorer}. Detailed spectral and temporal 
properties of the source are studied. Low and very low frequency quasi-periodic oscillations (QPOs), with a general 
trend of increasing frequency are observed during the outburst. Spectral analysis is done using the combined data of 
PCA and HEXTE instruments with two types of models: the well known phenomenological power-law model and the current
version of the {\it fits} file of two-component advective flow (TCAF) solution as an additive table model in XSPEC. 
During the entire period of the outburst, a non-thermal power-law component and the TCAF model fitted sub-Keplerian halo rate
were found to be highly dominant. We suggest that this so-called outburst is due to enhanced 
jet activity. Indeed, the `outburst' subsides when this activity disappears. We estimated X-ray fluxes coming from 
the base of the jet and found that the radio flux is correlated with this X-ray flux. 
Though the object was in the hard state in the entire episode, the spectrum becomes slightly softer 
with the rise in Keplerian disk rate in the late declining phase. We also estimated the probable mass of the source 
from our spectral analysis with the TCAF solution. Our estimated mass of XTE~J1118+480 is $6.99^{+0.50}_{-0.74}~M_\odot$ 
i.e., in the range of $6.25-7.49~M_\odot$.
\end{abstract}


\section{Introduction}
Black holes (BHs) are the extreme compact end result of massive stars. The gravitational attraction becomes 
so strong around them that even light cannot escape from inside the event horizon. Major inferences 
about these enigmatic objects are made from the radiation of the accretion of matter. 
The infalling matter radiates energy when swirling towards the compact object. 
Stellar mass BH sources which are normally 
in quiescent phase, become active in X-rays and $\gamma$-rays when they accrete significant amount of 
matter stored in accretion disks. These BH sources are known as the transient black hole candidate (BHC) 
and the phase of the sudden rise in their photon count rates is termed as the `outburst' phase. During an 
outburst phase, they show rapid variations in timing and spectral properties. There is an abundance of scientific literature, 
where extensive studies of the spectral and the temporal characteristics of several outbursting BHCs 
carried out till date have been reported (see for example, Tomsick et al. 2000; McClintock \& Remillard, 2006; 
Debnath et al., 2008; Nandi et al. 2012; Rao et al. 2013).

XTE J1118+480 is one such a Galactic transient low mass X-ray binary. This object was discovered 
by All Sky Monitor (ASM) onboard Rossi X-Ray Timing Explorer (RXTE) on 2000 March 29 at a sky location 
of R.A.=$11^h18^m10^s.79$, Dec.=$48^\circ02'12''.42$ in the Ursa Major constellation (Remillard et al. 2000).
During that time, this compact object was going through an apparent enhancements in X-rays, which may be termed as an outburst. 
Retrospective ASM analysis indicated that the source was actually discovered in its second outburst, the first 
outburst occurred during Jan 2-29, peaking at around $37$~mCrab on 2000 Jan 6 (Remillard et al., 2000).
Near-infrared, optical, extreme ultraviolet and radio observations followed 
(Uemura et al., 2000; Chaty et al., 2000; Wren \& McKay, 2000; Mauche et al., 2000; Pooley \& Waldram, 2000) after 
the announcement of its discovery. Very low frequency Quasi Periodic Oscillations (QPOs) were observed during 
its 2000 outburst (Wood et al. 2000). QPOs were detected in X-rays, EUV, and also in optical wavelength (Haswell et al. 2000).
Multi-wavelength studies were carried out for 2000 outburst by several authors (Cook et al., 2000; Garcia et al., 2000; 
Haswell et al., 2000; Hynes et al., 2000; McClintock et al., 2000; Wagner et al., 2000; Taranova et al., 2000; Esin et al., 2001;
Markoff et al., 2001; Hynes et al., 2003; Chaty et al., 2003). Revnivstev et al. (2000) argued that this compact object is a 
black hole due to the lack of high frequency variability in its power spectrum. From spectroscopic, 
photometric and dynamical analysis, Wagner et al. (2001) confirmed it as a Galactic halo black hole. 
Dynamical mass measurement for this object has been done by several workers: $6.0 - 7.7\ M_\odot$ (Wagner et al., 2001),
$8.53\pm0.6~M_\odot$ (Gelino et al., 2006) and $6.9-8.2 \ M_\odot$ (Khargharia et al., 2013)).
The companion of this compact object is a K5 spectral type star with $0.09 - 0.5$ $M_\odot$
(Wagner et al. 2001). The distance of this system is approximately 6000 light years 
or around 1.8 kpc (McClintock et al., 2001a). The estimated high galactic latitude of $62^\circ$ placed it in 
the Galactic halo (Uemura et al., 2000). The inclination of this compact object is estimated to be approximately 
$68-79^\circ$ (Khargharia et al., 2013). This short orbital period binary has an approximate $4.1$~hrs of orbital period 
(Patterson et al., 2000; Gonzalez Hernandez et al., 2012).

There are many phenomenological and theoretical models in the literature, with which one can study spectral and 
temporal properties of black hole sources. A two component advective flow (TCAF) model is a theoretical model 
which is based on transonic flow solution with a self-consistent radiative transfer and hydrodynamics. 
It was introduced by Chakrabarti and his collaborators in mid-90s (Chakrabarti, 1995; Chakrabarti \& Titarchuk, 1995; Chakrabarti, 1997). 
In TCAF paradigm, the accretion flow around a black hole consists of two components: high angular momentum, 
high viscous, optically thick and geometrically thin Keplerian disk, which produces soft multi-color blackbody part 
of the observed spectrum and another component is low angular momentum, low viscous, optically thin and geometrically thick 
sub-Keplerian matter. Keplerian disk moves on the equatorial plane, sandwiched by the sub-Keplerian advective flow.
The sub-Keplerian matter moves towards the black hole in almost free-fall time scale. Due to increasing centrifugal
barrier close to the black hole, matter which was moving super-sonically, slows down to become sub-sonic 
and as a result of that a shock is formed at a location depending on the energy and angular momentum of the 
sub-Keplerian flow. Due to conversion of the radial kinetic energy to thermal energy,
the post-shock region becomes `hot' and puffed-up. It is called the CENtrifugal pressure dominated BOundary Layer (CENBOL). 
CENBOL acts as so-called `hot-corona', which is responsible for the hard power-law part of the observed spectrum. 
The soft photons originating from the Keplerian disk get inverse Comptonized by the hot electrons of CENBOL and become hard. 
Some of these hard photons from the CENBOL get reflected in the Keplerian disk and modify the disk temperature.
The relative dominance of these two types of matter (Keplerian and sub-Keplerian) and shock parameters (location and strength) 
decide the spectral states of a black hole. High supply of the sub-Keplerian matter and comparatively low supply of the Keplerian 
matter make the spectrum hard, where power-law photons dominate. When the relative supply of the Keplerian matter increases, 
the CENBOL region starts to cool down and shrinks. As a result, the spectrum becomes softer. 
The intermediate states occur when these two rates become comparable to each other.
The oscillations of the shock boundary i.e., as a whole CENBOL cause quasi-periodic oscillations (QPOs). 
The oscillation of the shock occurs when resonance condition satisfies, i.e., when cooling and infall time scales 
roughly matches with each other (Molteni et al., 1996; Chakrabarti et al., 2015), or when Rankine-Hugoniot conditions 
do not satisfy to form a stable shock (Ryu et al., 1997).

The CENBOL is also believed to be the base of the jets or outflows.
The outflow rate depends on the accretion rate and shock strength (Chakrabarti, 1999a, 1999b). The outflow is maximum when
the shock has intermediate strength. Different types of jets are observed in different spectral states (Chakrabarti, 2001; 
Fender et al., 2004; Corbel et al., 2011; Coriat et al., 2011).
In the hard state, matter moves outward due to the pressure gradient force. Initially subsonic flow, after passing through sonic
surface, it becomes supersonic and moves away. The temperature falls due to
expansion and it emits in X-ray, UV, IR and radio in succession. A compact jet is often observed in hard states. In intermediate states, more
matter comes in and cools the CENBOL. Hence the matter near the sonic surface become supersonic and gets separated from the CENBOL.
It moves away as a blobby jet or discrete ejection. In the soft state, Keplerian disk completely cools down the CENBOL, hence jet
is quenched. There are other models of jet formation in the literature which invoke acceleration and collimation by magnetic fields 
Blandford \& Konigl, 1979; Blandfod \& Payne, 1982; Falcke \& Biermann, 1995; Markoff et al., 2001; Heinz \& Sunyaev, 2003)

Recently, after the implementation of this TCAF solution into HeaSARC's spectral analysis software package XSPEC
(Arnaud 1996) as an additive table model (Debnath et al. 2014), a clear picture of the physical flow properties in
several black hole candidates (e.g., H~1743-322, GX~339-4, MAXI~J1659-152, MAXI~J1836-194, MAXI~J1543-564, Swift~J1753.5-0127)
are obtained (see, Debnath, 2014, 2015a,b, 2017; Mondal et al., 2014, 2016; Jana et al., 2016;
Chatterjee et al., 2016; Molla et al., 2017; Bhattacharjee et al., 2017). From the TCAF model fitted
flow parameters, such as shock locations and strengths one can also predict frequency of the dominant QPOs (see, Debnath et al. 2014;
Chatterjee et al. 2016 and references therein). Unknown mass of an BHC could also be estimated from TCAF model fit (Molla et al. 2016).
The estimation of jet X-ray flux from spectral analysis with the TCAF model is also possible when the source is active 
in jet (see, Jana et al., 2017, 2018). Recent studies of a few transient BHCs with the TCAF solution, motivated us 
to use the same solution to study detailed spectral and temporal properties of XTE~J1118+480 during its 2000 outburst.

According to Pooley \& Waldram (2000), Dhawan et al. (2000), Fender et al. (2001), and Chaty et al. (2003), 
the source was highly active in radio with roughly constant flux of $\sim 9$~mJy from 2000 March to July. 
Chaty et al. (2003) also reported that radio flux at $15~GHz$ Ryle Telescope decreased to $\sim0.15~mJy$ 
in 2000 August. Many jet dominated models have been developed to explain the observed spectra 
(See, Markoff et al., 2001; Miller et al., 2002; Malzac et al., 2005; Maitra et al., 2009; 
Pe'er \& Markoff, 2012; Zhang \& Xie, 2013) Since the object was jet dominated, the X-ray emitted from the base of the jet is likely 
to contribute to enhance the overall X-ray flux. In this paper, we show that this is indeed the case.

The {\it paper} is organized in the following way. 
In \S 2, we briefly describe observation and data analysis techniques using HEASOFT software package. In \S 3, we present 
spectral and temporal analysis results of the source during its 2000 outburst. The estimation of X-ray 
fluxes coming from jets/outflows have been made. We also estimate the most probable range of the mass of the BH, which is an 
outcome of our spectral analysis using TCAF solution. Finally in \S 4, we present our view on the genesis of the so-called outburst in 2000. 

\section{Observation and Data Analysis}

After the discovery of the BHC XTE J1118+480 in March 2000, RXTE PCA monitored it roughly on a daily basis, 
till August 2000. We choose $32$ observations, from 2000 Mar 29 (MJD=51632.97) to Jul 30 (MJD=51755.08) 
to study evolution of the spectral and the temporal properties during this outburst. 
We use HEASARC's software package HeaSOFT version HEADAS 6.16 and XSPEC version 12.8 for analyzing 
the RXTE PCA data. For data reduction and analysis, we generally follow the methods as mentioned 
in Debnath et al. (2013, 2015a).

For timing analysis, RXTE PCA light curves in the energy range of $2-15$~keV and $2-25$~keV are generated using the 
event mode data of maximum time resolution of 125$\mu$s. The Power Density Spectra (PDS) are generated using XRONOS 
task `powspec' on $0.01$~sec binned light curves. Centroid frequencies of the observed QPOs (in PDS) are obtained   
from their fits with the Lorentzian profiles. To calculate average PCA count rate in $2-25$~keV for each observation, 
we use $1$~sec time binned background subtracted light curves of the proportional counter unit 2 (PCU2).

For the spectral analysis, PCU2 data of `standard 2' mode (FS4a* in the energy range of $3-25$~keV) and HEXTE science 
mode data (FS52* in the energy range $20-100$~keV) of Cluster 0 or A are used. The combined $3-100$~keV (except for 
the last three observations, where only PCU2 data in $3-25$~keV is used due to low signal-to-noise ratio in HEXTE band) 
background subtracted spectra are first fitted with the single non-thermal power-law (PL) model. Importantly, we do not 
see any significant contribution from the thermal disk blackbody component, while fitting the spectra with the PL component 
during the 2000 outburst. The whole part of the spectrum can be solely fitted with a single power-law component, 
no broken power-law needed either. We then refit all the spectra from the outburst with the current version (v0.3) 
of the TCAF model {\it fits} file. To fit BH spectra with 
the TCAF model in XSPEC, one requires four model input parameters, the mass of the BH and normalization. 
The normalization is assumed to be constant for a known BHC for a particular instrument (Molla et al. 2016, 2017). 

The model input parameters are: 
$i)$ Keplerian disk rate ($\dot{m_d}$ in Eddington rate $\dot{M}_{Edd}$), $ii)$ sub-Keplerian halo rate ($\dot{m_h}$ 
in $\dot{M}_{Edd}$), $iii)$ location of the shock ($X_s$ in Schwarzschild radius 
$r_s$=$2GM_{BH}/c^2$), $iv)$ compression ratio ($R=\rho_+/\rho_-$, where $\rho_+$ and $\rho_-$ 
are the densities in the post- and the pre- shock flows) of the shock. 
In case the mass of the BH ($M_{BH}$ measured in solar mass $M_\odot$) and normalization ($N$) 
are not known, they have to be treated as free parameters.   
In subsequent analysis, the derived values of the mass and normalization could be frozen in 
obtaining the fits. However, since the confidence level of the data are not the same for all observation 
days, the instrumental error may vary from day to day. Hence, for a fair scaling of the observed spectrum, 
we keep the mass parameter as a variable to incorporate these fluctuations. 
A fixed hydrogen column density ($N_H$) of $1.3\times10^{20}$ atoms cm$^{-2}$ 
for photoelectric absorption model {\it phabs} is used (Garcia et al. 2000, McClintock et al. 2001b). 
We also use a fixed value of 0.5\% of the systematic error throughout the outburst, while fitting the spectra 
with the PL or with the TCAF model. The positive and negative errors for the model fitted parameters 
are calculated using XSPEC command `err' after obtaining best fit, based on $\chi^2_{red} \simeq 1$.

Since base of the jets may also produce X-rays, we were particularly interested to see if the normalization is reduced 
when there is no jet activity. Recently Jana, Chakrabarti \& Debnath (2017; hereafter JCD17) developed a 
method to calculate X-ray flux contribution from the jets by treating the normalization during inactivity 
of jets to be the true normalization and higher normalization is simply due to the contribution from the 
jet. We shall present the flux of the estimated jets and its properties below and also show that the 
normalization is directly related to the radio flux.

\section{Results}

To study accretion flow properties of XTE J1118+480, we analyze archival data of RXTE PCA and HEXTE. 
$32$ observations from the 2000 outburst are selected for our timing and spectral analysis of the source. 
Spectra are fitted with two types of models: one is the phenomenological model, PL, and other is the 
physical model, namely, TCAF. PL model fits give us a rough estimation of non-thermal flux 
contributions without telling us about the exact physical cause, whereas the TCAF model fits provide us detailed picture 
of the accretion flows and associated physical processes around the BH. The combined spectral and temporal 
studies of the source during its 2000 outburst allow us to conclude that the source was in 
a hard state during the entire period of the outburst, as reported earlier in the literature (Hynes et al. 2000; 
Chaty et al. 2003). Moreover, the count rates are marginally higher 
(giving an impression of an outburst like behaviour) when the jet is active.

The detailed analysis results are noted in Tables 1, 2 \& 3. In Table 1, Lorentzian model fitted 
observed QPO frequency (dominating primary) values are mentioned. Power-law and TCAF model 
fitted spectral parameters are mentioned in Table 2. Contributions of total X-ray flux, 
X-ray flux from accretion disk and Jets are given in Table 3. Percentage contributions of 
Jet X-ray flux to the total X-ray flux are also shown. TCAF model fitted normalization values 
with errors are given in Col. 8.

Figures 1(a-d) show variations of PCA count rate, TCAF model fitted Keplerian disk rate ($\dot{m}_d$),
sub-Keplerian halo rate ($\dot{m}_h$) and $3-100$~keV PL model flux with MJDs, except for the last three 
observations, which are fitted in $3-25$~keV energy range. Variation of the mass of the BH $M_{BH}$ in $M_\odot$, 
shock locations ($X_s$) and compression ratios ($R$) with MJDs are shown in Fig. 2(a-c). In Fig. 2(d-e), 
variation of PL photon indices ($\Gamma$) and observed QPO frequencies with MJDs are shown. Figures 3(a) and 3(b) 
are the unfolded TCAF model fitted spectra for observation IDs 50133-01-01-00 (MJD=51642.57), 50407-01-07-00 (MJD=51683.50) 
respectively. In Fig. 4(a-b), variation of TCAF model fitted $\chi^2_{red}$ with different $M_{BH}$ grid values for 
above mentioned two spectra are shown. In Fig. 5(a-c), variation of total flux ($F_x$), inflow flux ($F_{inf}$), 
outflow flux ($F_{ouf}$) are shown respectively in units of $10^{-9}~erg~cm^{-2}~s^{-1}$. TCAF model fitted normalization 
variation with MJDs is shown in Fig. 5(d). In Fig. 5(e), radio data of $15.2~GHz$ are shown in units of $mJy$ 
(http://www.mrao.cam.ac.uk/$\sim$guy/J1118+480/J1118480.list). In Fig. 6, correlation plots of 3-25 keV outflow flux 
($F_{ouf}$)(green) and normalization (N)(red) with radio flux ($F_R$) are shown using quasi-simultaneous observations.

\subsection{Temporal properties}
Timing analysis with the PCA data shows that there is a slow rise in PCU2 photon count rate in $2-25$~keV energy band 
during the outburst for initial $\sim 26$~days, until 2000 Apr. 24 (MJD=51658.86). After that, the count rate decreases 
monotonically (see, Fig. 1a). The nature of the count rate variation shows that the source belongs to the
{\it slow-rise slow-decay} (SRSD; Debnath et al. 2010) class of objects, although this is really not a true outburst 
since the spectrum monotonically softens even towards the end of the so-called outburst.

Very low and low frequency QPOs ($0.06-0.16$~Hz) are observed for $13$ observations out of $28$ 
(see, Fig. 2d \& Table 1). Though we find sporadic behaviour of QPOs from only PCA observations, 
Wood et al. (2000) reported a consistent monotonically increasing nature of QPO frequencies while analyzing both 
RXTE (PCA) and USA satellites data. This could be due to low signal-to-noise ratio in the low frequency region. 
First prominent QPO of $0.06$~Hz is detected on the 2nd observation day (2000 Mar 31; MJD=51634.62). 
After that, observed QPO frequencies show a general trend of increasing nature up to 2000 June 15 
(MJD=51710.83), when a prominent QPO of $0.16$~Hz is observed. No significant QPO is found after MJD=51710.83, 
till the end of our study (MJD=51733.86). The disappearance of QPOs could be due to the non-satisfaction 
of resonance condition between cooling and infall time (compressional heating) scales (Molteni et al. 1996; Chakrabarti et al. 2015). 
The physical reasons behind the observation of these very low frequency QPOs and their slow movement 
during the outburst have been discussed more details in \S 4.

\subsection{Spectral properties}

RXTE PCA and HEXTE spectral data of the 2000 outburst of XTE J1118+480 in the $3-100$~keV energy band are fitted 
only with the PL model component. The last three observations are fitted in the $3-25$~keV energy range. 
While fitting spectra with the PL model, we have not seen any significant contribution in the disk component. 
So, we have not added extra disk black body component. The entire 2000 outburst was highly dominated by the 
non-thermal PL model component. During the initial phase of the outburst, PL photon index ($\Gamma$) remains nearly 
constant $\sim 1.72$, before its starts to increase slowly during the declining phase of the outburst (see, Fig 2d and Table 1). 
The PL model flux is also observed to vary with a similar trend of slow rise and slow decay as the variation 
of the $2-25$~keV PCA count rate (Fig. 1a \& 1d). 

To understand the physical picture of the accretion flow dynamics during the outburst, we refitted all PL fitted 
spectra with the current version of the TCAF model {\it fits} file. From each TCAF fitted spectrum, we get an 
estimation of the size (location), density, temperature, etc. of the `hot' Compton Cloud (CENBOL), probable 
mass of the black hole, other than the values of the inflowing Keplerian disk and sub-Keplerian halo 
mass accretion rates. The halo rate ($\dot{m}_h$) varies (Fig. 1c) roughly in a similar way as the PCA count rate
during the outburst (Fig. 1a). On the first observation day (MJD=51632.97), the halo rate is found at 
$0.336$~$\dot{M}_{Edd}$, and then it starts to increase and reaches a maximum value of $0.353$~$\dot{M}_{Edd}$ 
on 2000 Apr 24 (MJD=51658.86). After that for the next four observations, until MJD=51668.22 $\dot{m}_h$ decreases 
faster (moves to $0.321$~$\dot{M}_{Edd}$), then decreasing-rate of $\dot{m}_h$ becomes very slow until MJD=51719.92 
($0.316$~$\dot{M}_{Edd}$). After that, within $\sim 31$ days it decreases to $0.284$~$\dot{M}_{Edd}$ at the 
last observation (MJD=51755.08).

During the entire period of the outburst, very low values ($0.018-0.044~\dot{M}_{Edd}$) of the Keplerian disk rates ($\dot{m}_d$) 
are required to fit the spectra, which clearly indicate the dominance of the low angular momentum halo matter over the high viscous 
disk component (Fig. 1b \& c). We also notice that starting from the 2nd observation $\dot{m}_d$ as well as $\dot{m}_h$, are almost 
constant until 2000 April 24 (MJD=51658.86). On this observation maximum values of PCA count rate, PL flux and $\dot{m}_h$ are observed. 
After that until the end of the outburst, $\dot{m}_d$ increases slowly (from $0.021$ to $0.044~\dot{M}_{Edd}$) whereas 
$\dot{m}_h$ decreases slowly (from $0.353$ to $0.284\dot{M}_{Edd}$, Fig. 1 and Table 1). As a result 
of this, after MJD=51658.86, the spectrum becomes slightly softer (within the hard spectral band) and it is evident also from the slow rise 
in PL photon indices (see, Fig. 2d). This shows that this object is not undergoing an outburst in the conventional sense. The physical 
explanation of these behaviours of the source during the outburst has been explained more details in \S 4.

As the outburst progresses, at first, the shock moves slightly inward from $X_s=387~r_s$ on the 2nd observation day (MJD= 51634.62) 
to $X_s= 383~r_s$ (MJD= 51649.33; see Fig. 2b). Then it begins to shift outward for the next two observations (MJD= 51652.54 \& 51655.18) 
and again starts to move inward after this day. The location of the shock changes randomly within a small range from $X_s=379$ to $X_s=428~r_s$ 
during entire period of the outburst. On MJD=51738.71, $X_s$ becomes maximum ($X_s=428~r_s$) and then $\sim$9~days (MJD=51747.81) 
it moves inward until $X_s=380~r_s$. After that, $X_s$ roughly remains constant until our last observation.
The value of shock compression ratio ($R$) is $3.394$ on the first observation day (MJD=51634.62). 
Then $R$ increases very slowly from the second observation day and becomes maximum on MJD=51706.09 ($R=3.855$), 
then it starts to decrease (Fig. 2c). However, the range of variation of $R$ during the entire 2000 outburst is very small ($\sim~3.269-3.855$).
This behaviour is not similar to any of the outbursts we analyzed before and thus we do not believe that the present event is
an outburst in an conventional sense.

Based on the variation of the TCAF model fitted physical accretion flow parameters ($\dot{m}_d$, $\dot{m}_h$, $X_s$, $R$), 
nature of QPOs and PL photon indices ($\Gamma$), we have come to the conclusion that during entire period of the 2000 outburst, XTE J1118+480 
was in low-hard / hard state (HS). The non-observation of intermediate and soft spectral states could be due to the fact that 
outburst was not triggered by any viscous mechanism, but due to the X-rays from jets.

\subsection{Black Hole Mass Estimation}

Each TCAF model fit gives us one best fitted mass (M$_{BH}$) value with some uncertainties, 
since here mass of the BH is an input parameter to fit the spectrum. We get an average value of the mass 
of XTE~J118+480 as $6.99^{+0.50}_{-0.59}~M_\odot$ from the overall spectral fitting.

We apply another technique to obtain the probable mass of the BH from each the best fitted spectrum. Here, we use 
$M_{BH}$ $vs.$ $\chi^2_{red}$ method as discussed in Molla et al. (2016) and Chatterjee et al. (2016). We vary 
TCAF model input $M_{BH}$ value into different `+'ve and `-'ve fixed (frozen) grid points from its best-fitted 
values to observe model fitted $\chi^2_{red}$ values for each observations. Then variations of the $\chi^2_{red}$ 
with the $M_{BH}$ grid values are plotted (see, Fig. 4). 
We draw a horizontal line at 90\% confidence ($\chi^2_{red}=2.7$) for acceptable limit of the model fits, 
and two intercepting points give us $\pm$ limiting $M_{BH}$ value of each best fitted spectrum.

In Fig. 3(a-b), we show two best-fitted spectra (when all model parameters are kept free) of observation 
IDs: 50133-01-01-00 (MJD=51642.57) and 50407-01-07-00 (MJD=51683.50), selected from two different parts 
of the light curve. The best fitted $\chi^2_{red}$ values are marked in the plots. The model fitted variation 
of the $\chi^2_{red}$ values with $M_{BH}$ of these two spectra are shown in Fig. 4(a-b) respectively. 
Up to allowed 90\% confidence $\chi^2_{red}$ value, we obtain mass ranges for these two spectra as 
$6.78-7.13~M_\odot$ and $6.89-7.39~M_\odot$ respectively. A similar analysis is made for all observed 
spectra and finally we obtain the mass range of the source to be $6.25-7.40~M_\odot$.

Now combining the results of above mentioned methods, we finally estimate the probable mass range of the source 
as $6.25-7.49~M_\odot$ or $6.99^{+0.50}_{-0.74}$. 

\subsection{Estimation of X-ray Flux Contribution from Jets}

Recently JCD17 developed a method to calculate X-ray flux contribution from jets. 
They used the property of the constant normalization of the TCAF model (see also, Molla et al. 2016; 2017). 
If the jet is active then higher values of model normalization, since X-rays from the base of the jet is not
not included in the model {\it fits} file of the current version of the TCAF. 

While fitting combined PCA and HEXTE spectra of XTE J1118+480, TCAF model normalization ($N$) is observed 
to vary in a narrow range of $\sim 14-18$ from 2000 Mar 29 (MJD=51632.97) to 2000 Jun 28 (MJD=51723.18). 
After that it starts to decrease with time and reaches its minimum value, $N=4.369$ on the our last observation, 
i.e., on 2000 July 30 (MJD=51755.08). In Fig. 5e, we show variation of the radio flux of $15~GHz$ Ryle Telescope 
in between 2000 Mar 30 (flux=$6.0$~mJy) to 2000 Jun 21 (flux=$10.15$~mJy). It seems that variation of our model 
normalization is consistent with the variation of the observed radio fluxes. Model normalization rapidly decreases 
after 2000 Jun 28, where no radio observations are reported. According to Chaty et al. (2003), radio flux was 
decreased to very low value in Aug., 2000. So, according to JCD17, one can say that most of the time (starting 
from very first day), the source is highly jet dominated and as the day progresses, specially after 
2000 Jun 28 (MJD=51723.18), X-ray contribution from jets decreases and becomes lowest or negligible on our 
last observation (on 2000 Jul 30; MJD=51755.08), when the minimum value of $N$ is observed. So, we suggest that 
only the accretion disk contributes to X-rays on that observation.  

For measuring of jet fluxes, we only use RXTE PCA data in $3-25$~keV, since in the last three observations, we are
unable to get good S/N data in the HEXTE band. To calculate flux in $3-25$~keV, we use XSPEC command `flux 3.0 25.0' 
after obtaining best model fits. Total X-ray fluxes ($F_X$) are calculated when all TCAF model parameters are kept free. 
It gives us contribution of X-rays coming from both inflowing matter or accretion disk and from outflowing matter or jets. 
To calculate X-ray flux contributions coming only from inflowing matter or accretion disk ($F_{inf}$), we refit 
all the spectra by keeping model normalization frozen at previously fitted lowest normalization value i.e., at $N=4.369$. 
Now, by taking differences the accretion disk X-ray flux ($F_{inf}$) from total X-ray flux ($F_X$), one can get 
the jet X-ray flux ($F_{ouf}$),

$$F_{ouf} = F_X - F_{inf}  \eqno(1)$$

In Fig. 5, we show the variation of (a) $F_X$, (b) $F_{inf}$ and (c) $F_{ouf}$ in unit of $10^{-9}~ergs~cm^{-2}~s^{-1}$. 
In Fig. 5d, we also show the variation of the TCAF model normalization during entire outburst, when all model parameters are 
kept free. On the first observation day, $F_{ouf}$ is observed at $0.802\times10^{-9}~ergs~cm^{-2}~s^{-1}$. Then it starts to 
increase and attains its maximum value on 2000 Apr 24 (MJD=51658.86). Then $F_{ouf}$ starts to decrease very slowly. 
On 2000 Jul 22 (MJD=51747.81), it drops to a very low value and on the last two observations (2000 Jul 25 \& 30), $F_{ouf}$ 
becomes negligible. 

$F_R$ also shows a similar behaviour (see, Fig. 5e). On our second observation day, i.e., 2000 Mar 31 (MJD=51634.87), it was at $6$~mJy. 
After that, it increases and remains almost constant for the rest of the outburst at $\sim 9~mJy$. $F_{inf}$ starts to increase 
from the beginning of the outburst and attains its maximum value on 2000 Apr 29 (MJD=51663.20). Then it decreases a little bit 
and remains almost constant. For the last three days it drops to a lower value of $\sim 0.201\times10^{-9}~ergs~cm^{-2}~s^{-1}$.

For a quantitative study, we plot (Fig. 6, online green circles) the variation of X-rays obtained from the base of the jet 
($F_{ouf}$ in $3-25~keV$) with the radio flux ($F_R$). Though, $F_{ouf}$ comes from our spectral fit of X-ray data and separation 
of components (see, Eqn. 1) and $F_R$ are the quasi-simultaneous radio observations, we see a clear correlation between them. 
We fit the variation of $F_{ouf}$ and $F_R$ using the relation $F_R \sim F_{ouf}^b$, where $b$ is a constant. We obtain $b \sim 1.51 \pm 0.29$ 
(online green line). In Fig. 6, we also show the variation of the TCAF model normalization (N) with the observed radio flux (see, online red diamond points).
In this case, we obtain $b \sim 1.32 \pm 0.45$ (online red line). We see that both the X-rays from the outflow ($F_{ouf}$) and normalization
are well correlated with the radio ($F_R$) flux. The Pearson's Linear correlation coefficients ($LP_e$) for $F_{ouf}~vs.~F_{R}$ and 
$N~vs.~F_{R}$ are $\sim0.82$ and $\sim0.65$ respectively. These values suggest that $F_{ouf}$ and $N$ are strongly correlated with $F_R$. 
To confirm this, we also calculate Spearman's rank correlation coefficients ($SR_e$) for those two types of variations and obtain $SR_e$ 
as $\sim0.79$ and $\sim0.57$ respectively. 

\section{Discussion and Concluding Remarks}

A detailed temporal and spectral studies of the Galactic transient BHC XTE~J1118+480 during its very first `outburst' 
after its discovery on 2000 Mar 29 is done. 
The $3-100$~keV combined PCU2 and HEXTE spectra (except for the last three observations, where PCU2 data in $3-25$~keV 
band are used) are fitted with two types of models: PL and TCAF. While fitting the spectra with the phenomenological PL model, 
we do not find any significant blackbody contribution, probably because we use only RXTE PCA data having the lowest 
energy of $3$~keV. McClintock et al. (2001b) and Frontera et al. (2003) have found weak signatures of disk black body 
component since they have used low energy Chandra and BeppoSAX data. We presented evidence of a similar signature 
of low contribution of Keplerian disk matter as compared to the sub-Keplerian halo matter while spectra are fitted with
the TCAF model {\it fits} file. 

During the entire 2000 outburst, we see a roughly similar variation of the PCU2 rate (in $2-25$~keV), PL model flux 
and TCAF model fitted halo rate ($\dot{m}_h$). TCAF model fitted disk rate ($\dot{m}_d$) shows a general increasing 
trend (from $0.018$ to $0.044~\dot{M}_{Edd}$) during the outburst $0.018-0.044~\dot{M}_{Edd}$. The halo rate
remains roughly constant (in between $0.346-0.353~\dot{M}_{Edd}$) till MJD=51658.86, then it decreases for the next 
$\sim$ 20 days (from $0.353$ to $0.324~\dot{M}_{Edd}$). The decreasing rate slows down for a longer period 
(MJD=51672.46 to 51738.71). After MJD=51738.71, it starts to decrease rapidly for the last three observations. 
The PL model fitted photon indices also remain roughly constant ($\sim 1.72$) during initial phase of the outburst, 
before it starts to increase slowly during the declining phase of the outburst. 

Very low and low frequency QPOs are observed during the outburst. First significant QPO at $0.06$~Hz is 
observed on the second observation (MJD=51634.62) and after that it slowly increases up to MJD=51710.83. 
The maximum QPO frequency of $0.16$~Hz is observed on the last QPO observation day (MJD=51710.83). 
Although we have not found significant signatures of QPOs on a daily basis during MJD=$51634.62-51710.83$ 
with RXTE PCA data, Wood et al. (2000) observed monotonic evolution of the QPOs using combined PCA and USA satellite data. 
This could be due to low S/N PCA data in the low frequency USA satellite observed QPO region.
According to shock oscillation model (Molteni et al. 1996; Ryu et al. 1997), the frequency of the QPO is 
inversely proportional to the infall time. Since, during the current outburst of XTE~J1118+480, mHz QPOs are 
observed, shock should be far away from the BH event horizon. Although from our TCAF model fitted spectral analysis, 
we see that stronger shocks are formed at a location far away from the event horizon (see, Fig. 2b). However,  even 
these shock parameters (shock location and compression ratio) do not reproduce the QPO frequencies (see, Eqn. 2 of 
Chakrabarti et al. 2008) to explain these very low frequency QPO values. We suspect that these very low frequency 
QPOs are of non-conventional origin such as due to instabilities at the base of the jet. 
Miller et al. (2002) also suggested jet as the origin of these QPOs.
 
Depending upon the nature of the evolution of the TCAF model fitted physical accretion flow parameters 
($\dot{m}_d$, $\dot{m}_h$, $X_s$, $R$), PL photon indices ($\Gamma$) and nature of observed QPOs frequencies, 
we have come to the conclusion that the source was in HS during entire epoch of the outburst, which is quite 
unusual in outbursting BHCs. This appears to be due to dominance of X-rays coming from the jet, which is common 
in hard states. Indeed, we find that the intensity of X-rays die down when the jet activity is reduced. These unique 
and unnatural behaviour of the outburst has not been reported by any earlier published 
papers. The physical reason behind this nature could be dominance of the high magnetic field, which slowed down 
movement of the Keplerian disk (generally it moves in viscous time scale), which is evident from high radio flux 
during the outburst. The presence of high jet activity (evident from high radio flux) during most of the period 
of the outburst also supports our statement of high dominance of magnetic field.  

Recent studies showed that one could estimate mass of unknown BHCs quite successfully from spectral analysis 
with the current version of the TCAF model {\it fits} file (Molla et al. 2016, 2017; Chatterjee et al. 2016; 
Jana et al. 2016; Debnath et al. 2017). In this paper also, we estimated the most probable range of the mass of the Galactic 
BHC XTE~J1118+480 from our spectral analysis. This estimated mass value is verified using another method, 
namely minimum $\chi^2_{red}$ method. Each TCAF model fit provides one best fitted mass (M$_{BH}$) value with some uncertainties.
Mass was observed to vary in a narrow ranges of $6.40-7.49~M_\odot$ (see, Fig. 2a) when all model parameters are kept free. The average 
value of these best fitted mass is found to be $6.99^{+0.50}_{-0.59}~M_\odot$. From the $\chi^2_{red}$ method, 
we obtain the mass range of the source as $6.25-7.40~M_\odot$, which is almost consistent with the mass range obtained
from the spectral fitting. So, combining masses obtained from above two methods, the probable mass range 
of the source is found to be $6.25-7.49~M_\odot$ or $6.99^{+0.50}_{-0.74}$ M$_\odot$. This generally agrees with the mass estimate of others.
The estimated mass ranges from dynamical measurements are $6.0 - 7.7\ M_\odot$ (Wagner et al. 2001), $6.9-8.2 \ M_\odot$ 
(Khargharia et al. 2013). McClintock et al. (2001a) have prescribed the mass to be $\leq10~M_\odot$ from the I-band lightcurve variation. 
Gelino et al. (2006) has estimated the mass to be a slightly higher value ($M_{BH}$=$8.53\pm0.6~M_\odot$) from the simultaneous modelling 
of B, V, R, J, H, and K wavebands in quiescence state. They have also derived inclination angle as $68\pm2~^\circ$, 
which is lower than the value reported by others (for example, Wagner et al., 2001). This may be due to their assumption of 
negligible non-stellar flux contribution in Near-Infrared band (Wagner et al., 2001, Zurita et al., 2002, Khargharia et al., 2013). 
Our estimated mass is consistent with the value determined by Wagner et al. (2001), McClintock et al. (2001a) and Khargharia et al. (2013). 

We have also estimated jet X-ray fluxes from recently introduced method by JCD17, based on deviation of the constancy 
of the TCAF model normalization during the outburst. We separated the total X-ray flux ($F_X$) in $3-25$~keV PCA band 
into two components: one is coming from inflowing matter or from accretion disk ($F_{inf}$) and another is from outflowing 
matter or jets ($F_{ouf}$). Similar to model normalization ($\sim14-18$), our estimated $F_{ouf}$ is also observed 
in a narrow range ($\sim 0.8-1.2$) during the initial phase of the outburst until 2000 Jun 28 (MJD=51723.18), after 
that it starts to decrease and becomes negligible in the last few observations. 
During the high jet dominated region (2000 Mar 29 to Jun 28), average percentage contribution of the jet X-ray is 
found to be $\sim70\%$, with a maximum observed contribution of $\sim 75\%$. 
Although $F_{ouf}$ and normalization ($N$) are measured from spectral analysis, we tried to find correlation 
of $F_R$ with $F_{ouf}$ and $N$ (see Fig. 6). They are found well correlated. The Pearson's Linear (LP) and Spearman's rank (SR) 
correlation coefficients ($\sim 0.6-0.8$) also suggest that $F_{R}$ is strongly correlated with both $F_{ouf}$ and $N$.

Fender et al. (2001) have studied low/hard X-ray state of XTE~J118+480 in radio and sub-millimetre bands with VLA, Ryle Telescope,
MERLIN, and JCMT. They have concluded that the flat or inverted radio spectra are a general characteristic of low/hard state
and coming from the partially self-absorbed jet due to synchrotron emission. They have also inferred that a high amount of accretion 
power is extracted due to the outflow. This indeed support our conclusion of the jet activity induced `outburst'.
However, determination of the physical processes (i.e., synchrotron or Comptonization or both) associated with the X-rays from jet
is beyond the scope of the present work. Markoff et al. (2001) have constructed a model to explain the spectra of XTE J1118+480 
by including truncated disk and ADAF (also see, Miller et al., 2002; Malzac et al., 2005). They considered two processes such as 
synchrotron and inverse Comptonization to explain the spectra. For synchrotron process they assume the inner region 
of truncated disk to be around $100-1000~r_s$, which is quite similar to the TCAF fitted high $X_s$ values (see, Col. 9 of Table 2).

\section*{Acknowledgements}
D.C. and D.D. acknowledge support from DST/SERB sponsored Extra Mural Research project (EMR/2016/003918). 
A.J. and D.D. acknowledge support from DST/GITA sponsored India-Taiwan collaborative project (GITA/DST/TWN/P-76/2017) fund.




\clearpage

\begin{figure}
\vskip 0.8cm
        \centerline{
        \includegraphics[scale=0.6,width=8truecm,angle=0]{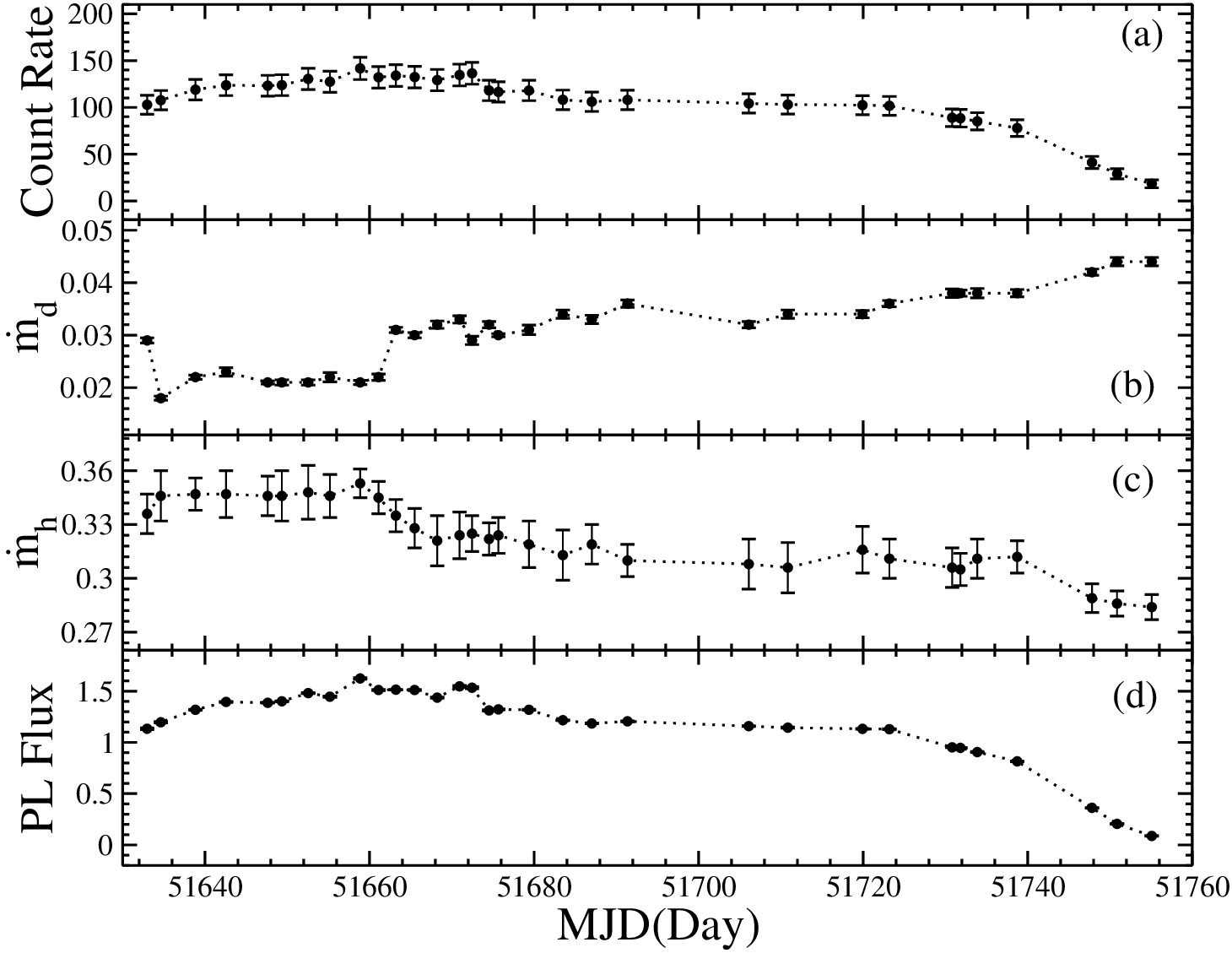}
        }
\caption{Variation of (a) PCA count rate (cnt/s) for 2-25 keV energy range, (b) disk rate ($\dot{m}_d$) in
        $\dot{M}_{Edd}$ (c) halo rate ($\dot{m}_h$) in $\dot{M}_{Edd}$, and (d) PL flux in unit of
        $10^{-9}~erg~cm^{-2}s^{-1}$ with day (MJD) are plotted.}
       \label{fig1}
\end{figure}

\begin{figure}
\vskip 0.8cm
        \centerline{
        \includegraphics[scale=0.6,width=8truecm,angle=0]{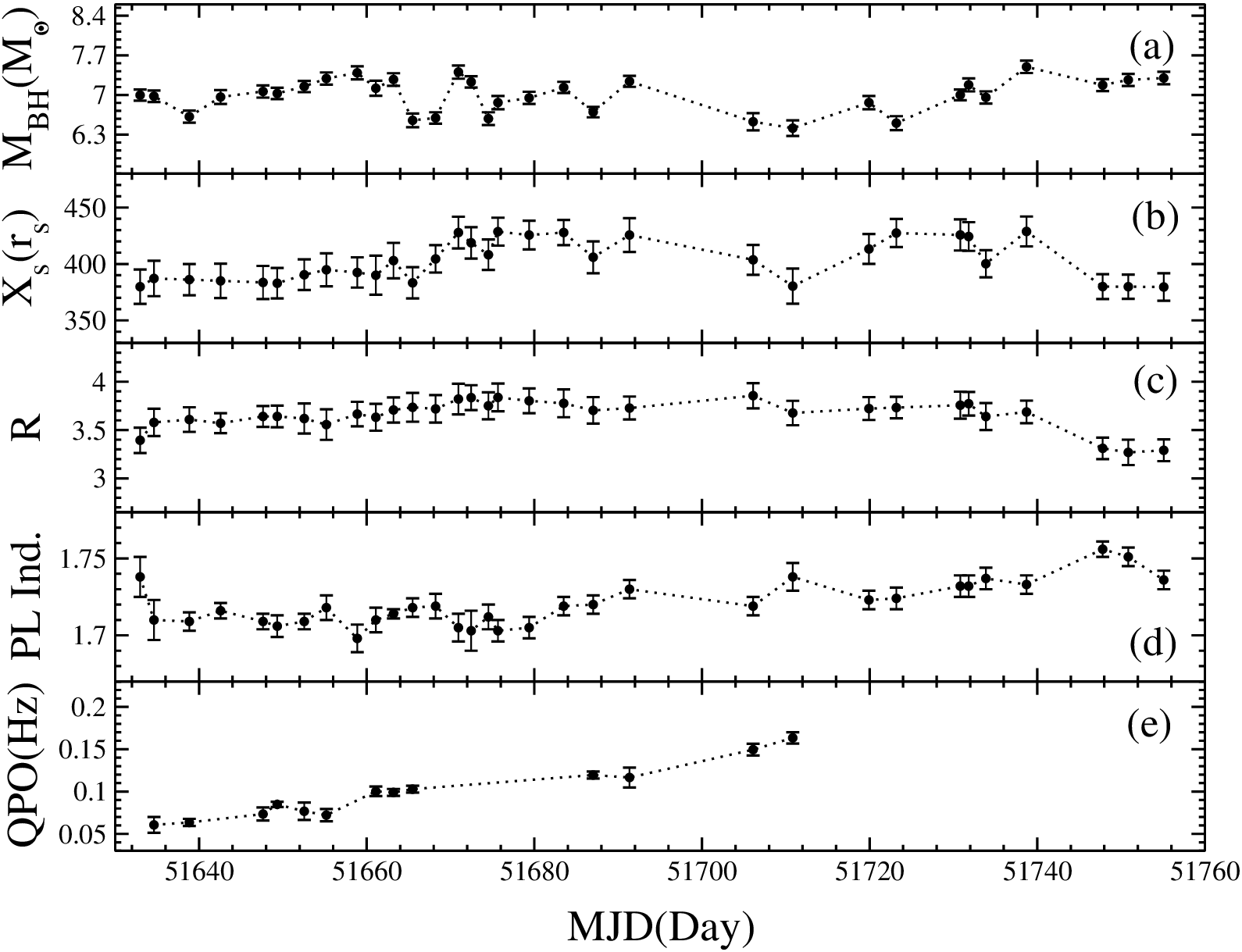}
        }
\caption{Variation of (a) mass of the BH, (b) shock location ($X_s$) in $r_s$, (c) compression ratio ($R$), (d) PL photon index ($\Gamma$),
         and (e) observed dominating QPO frequency (in Hz) with day (MJD) are shown. }
       \label{fig2}
\end{figure}

\begin{figure}
\vskip 0.8cm
        \centerline{
        \includegraphics[scale=0.6,width=6truecm,angle=270]{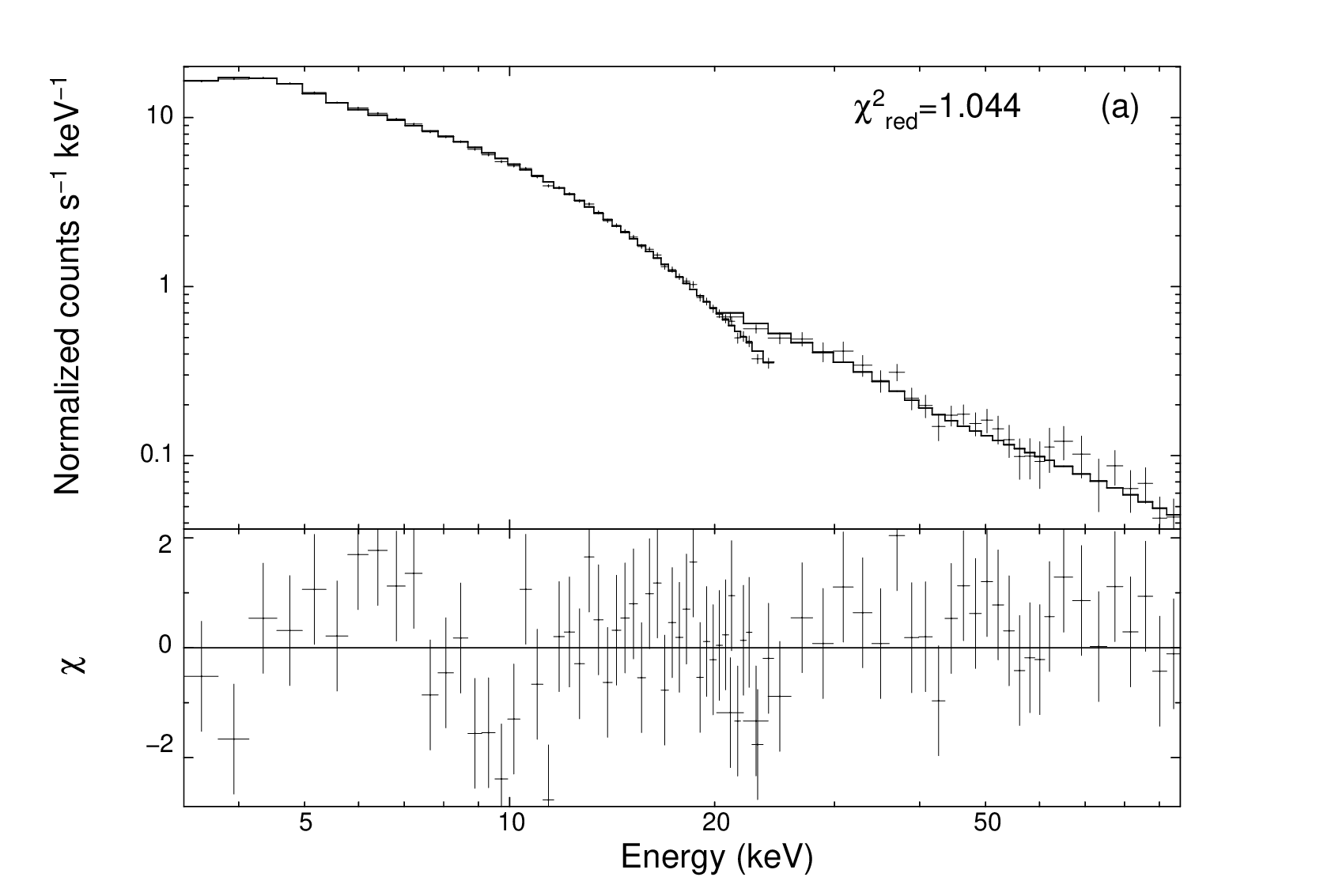}}\vskip 0.1cm
        \centerline{
        \includegraphics[scale=0.6,width=6truecm,angle=270]{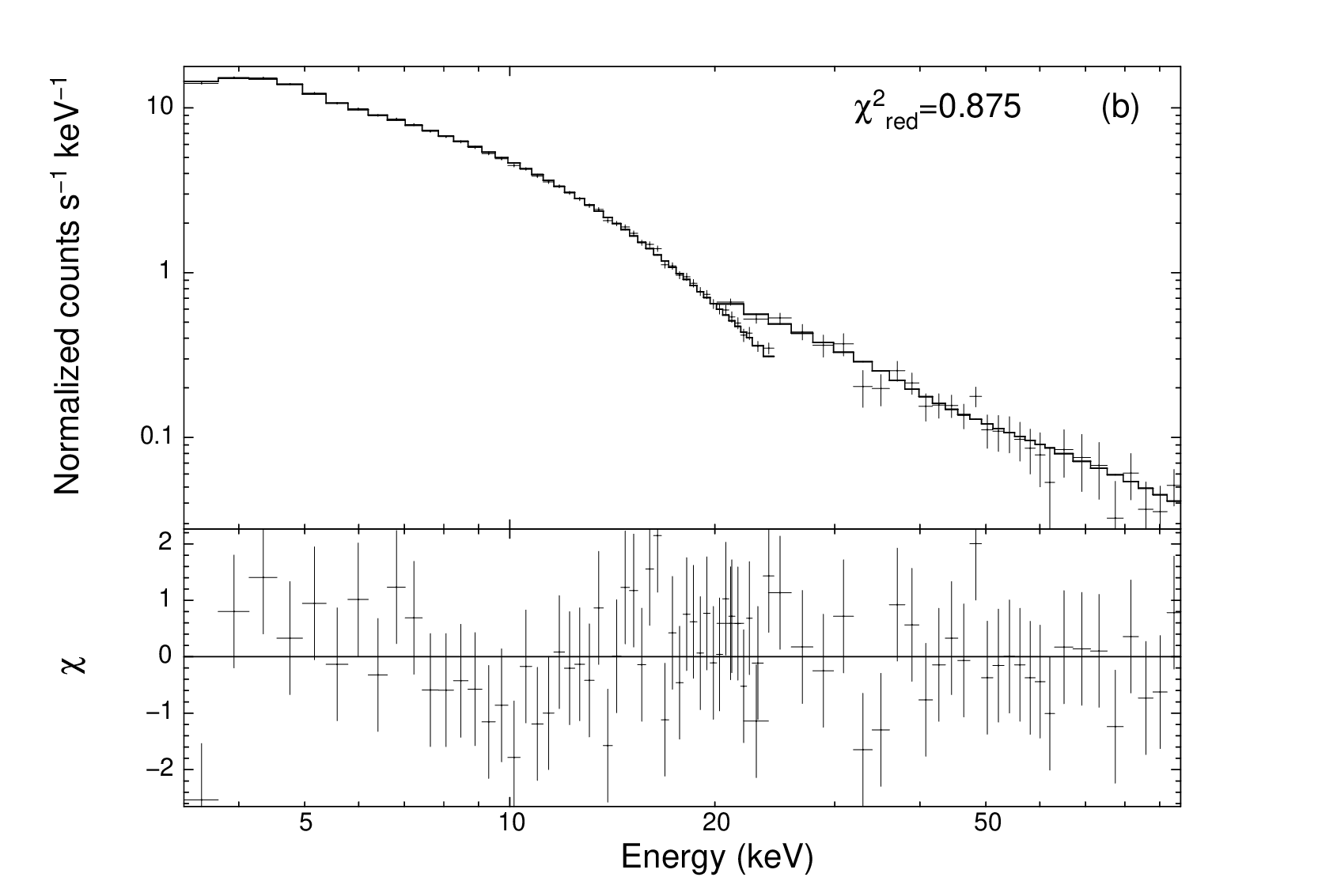}
        }
\caption{Normalized TCAF model fitted spectra for observation IDs (a) 50133-01-01-00, and (b) 50407-01-07-00 are shown.}
       \label{fig3}
\end{figure}
\begin{figure}
\vskip 0.8cm
        \centerline{
        \includegraphics[scale=0.6,width=9.0truecm,angle=0]{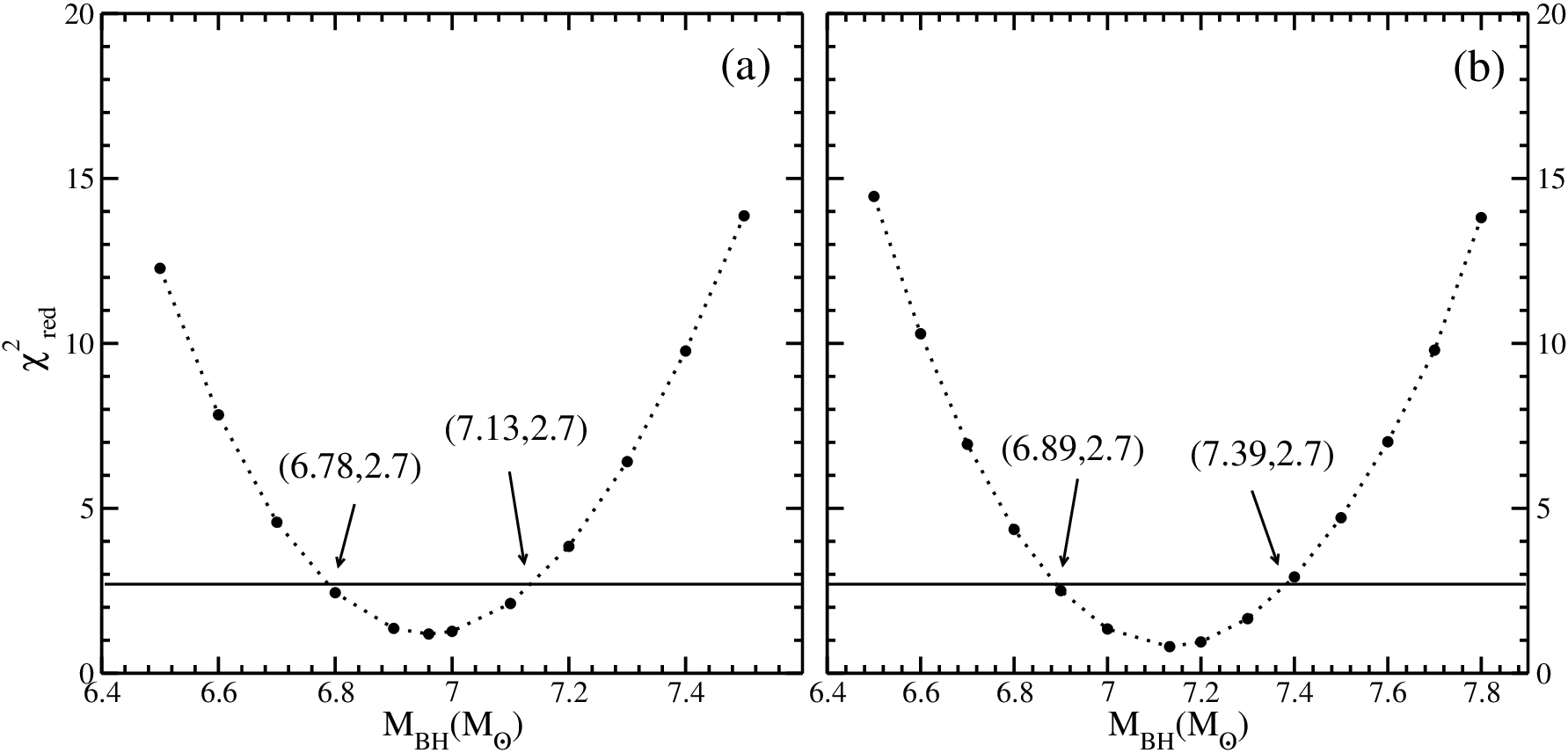}
        }
\caption{Variation of TCAF model fitted $\chi^2_{red}$ with mass for observation IDs (a) 50133-01-01-00, and (b) 50407-01-07-00
are shown. To see these variations we kept all TCAF model parameters as frozen into their best fitted values, expect mass of
the BH into different constant grid values as shown in the plots.}
       \label{fig4}
\end{figure}

\begin{figure}
\vskip 0.8cm
        \centerline{
        \includegraphics[scale=0.6,width=8truecm,angle=0]{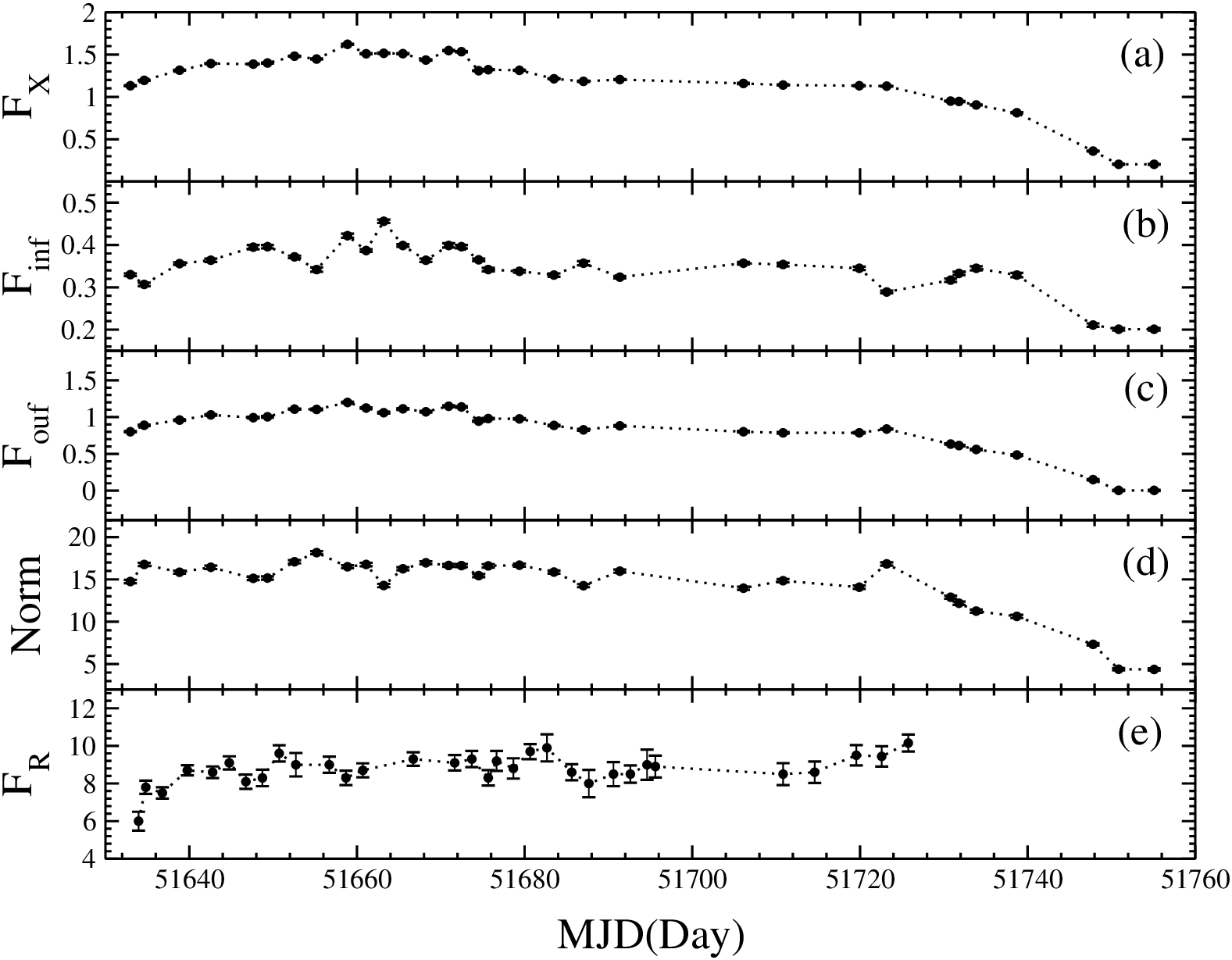}
        }
\caption{Variation of (a) Total flux, (b) inflow or disk flux, (c) outflow or jet flux, 
(d) TCAF model normalization and (e) radio flux are shown with MJD. $F_x$, $F_{inf}$, $F_{ouf}$
are in $10^{-9}~erg~cm^{-2}~s^{-1}$ unit. Radio flux of $15~GHz$ Ryle Telescope, is in $mJy$ unit 
and adopted from http://www.mrao.cam.ac.uk/$\sim$guy/J1118+480/J1118480.list.}
       \label{fig5}
\end{figure}

\begin{figure}
\vskip 0.8cm
        \centerline{
        \includegraphics[scale=0.6,width=8truecm,angle=0]{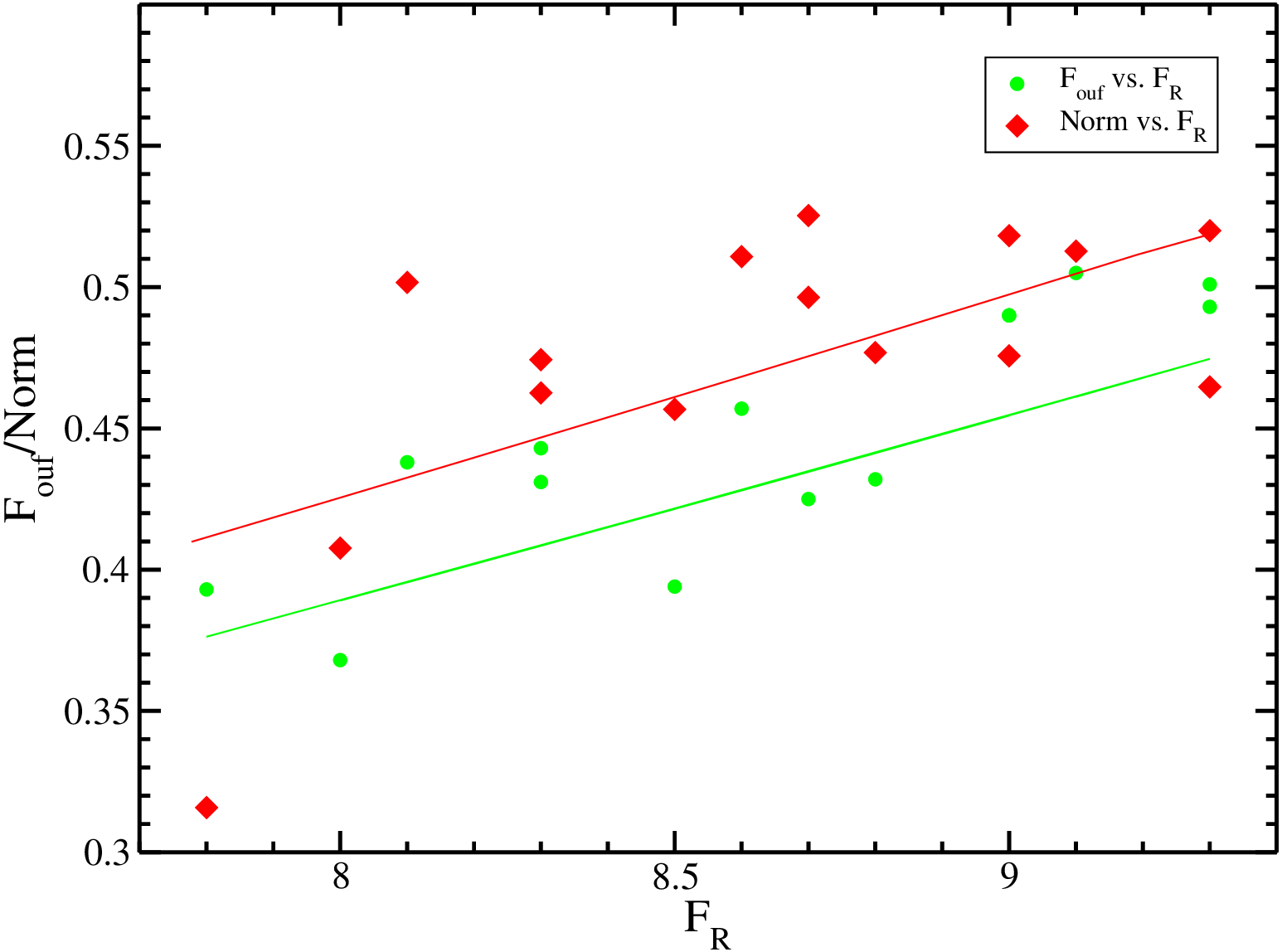}
        }
\caption{Correlation plots of $3-25$~keV jet X-ray flux ($F_{ouf}$; online green circles) and normalization 
(N; online red diamonds) with radio flux ($F_R$) of $15$~GHz~Ryle~Telescope are shown using quasi-simultaneous 
observations. The slope $b$ of $F_R \sim F_{ouf}^b$ (online green line) or $F_R \sim N^b$ (online red line) are 
found at $1.51 \pm 0.29$ and $1.32 \pm 0.45$ respectively.}
       \label{fig6}
\end{figure}

\begin{table}
\addtolength{\tabcolsep}{-3.0pt}
\centering
\caption{Observed QPO fitted parameters}
\label{tab:table1}
\begin{tabular}{lcccccc} 
\hline
Obs. ID & UT   & Day       &  QPO Freq.         & FWHM                 & Q & RMS  \\
        & Date & (MJD)     &  ($\nu$ in Hz)     &  ($\Delta\nu$ in Hz) &   & Amp.  \\
 (1)    &  (2) &    (3)    &       (4)          &   (5)                &(6)& (7)   \\
\hline
X-02-00 & 31/03 & 51634.62 & $0.061^{\pm 0.009}$ & $0.043^{\pm0.038}$ &1.412&2.074\\
X-05-00 & 04/04 & 51638.84 & $0.064^{\pm 0.004}$ & $0.032^{\pm0.010}$ &1.981&2.051\\
Z-01-01 & 13/04 & 51647.62 & $0.073^{\pm 0.007}$ & $0.046^{\pm0.030}$ &1.597&1.927\\
Z-02-00 & 15/04 & 51649.33 & $0.085^{\pm 0.003}$ & $0.025^{\pm0.013}$ &3.346&1.678\\
Y-02-00 & 18/04 & 51652.54 & $0.077^{\pm 0.010}$ & $0.071^{\pm0.021}$ &1.083&2.239\\
Z-03-00 & 21/04 & 51655.18 & $0.072^{\pm 0.007}$ & $0.034^{\pm0.016}$ &2.136&1.930\\
Z-03-02 & 27/04 & 51661.09 & $0.100^{\pm 0.005}$ & $0.033^{\pm0.021}$ &3.058&1.551\\
Y-03-00 & 29/04 & 51663.20 & $0.099^{\pm 0.004}$ & $0.040^{\pm0.020}$ &2.494&1.688\\
Z-04-02 & 01/05 & 51665.48 & $0.103^{\pm 0.004}$ & $0.037^{\pm0.018}$ &2.814&1.653\\
Z-07-01 & 23/05 & 51687.01 & $0.120^{\pm 0.004}$ & $0.039^{\pm0.013}$ &3.101&1.561\\
Z-08-00 & 27/05 & 51691.35 & $0.116^{\pm 0.011}$ & $0.061^{\pm0.034}$ &1.917&2.027\\
Z-10-00 & 11/06 & 51706.09 & $0.150^{\pm 0.006}$ & $0.071^{\pm0.023}$ &2.122&1.798\\
Z-10-01 & 15/06 & 51710.83 & $0.163^{\pm 0.006}$ & $0.041^{\pm0.020}$ &4.015&1.263\\
\hline
\end{tabular}
\noindent{
\leftline{Col. 1 represents observation IDs with prefix X=50137-01, }
\leftline{Y=50133-01, Z=50407-01. Columns 2 \& 3 show the date of } 
\leftline{observation in dd/mm format of year 2000, and days in MJD.}
\leftline{Cols. 4 \& 5 represent Lorentzian model fitted centroid frequency } 
\leftline{and full-width at half maximum (FWHM) of the observed QPOs in Hz.}
\leftline{Cols. 6 \& 7 represent `Q (=$\nu/\Delta\nu$)' and RMS amplitude of the QPOs.}
}
\end{table}

\clearpage
\begin{table}
\small
\addtolength{\tabcolsep}{-3.0pt}
\caption{PL and TCAF model fitted spectral parameters}
\label{tab:table2}
\begin{tabular}{lcc|ccc|cccccc}
\hline
Obs. ID& Date & MJD & PL Ind       & PL flux & $\chi^2/dof$ & $\dot{m}_d$       & $\dot{m}_h$       & $X_s$   & R  &  $M_{BH}$  & $\chi^2/dof$ \\
       &      &   &  ($\Gamma$)  &         &              & ($\dot{M}_{Edd}$) & ($\dot{M}_{Edd}$) & $(r_s)$ &    &    ($M_\odot$)&            \\
 (1)   & (2) &   (3)        &  (4)    &   (5)        &  (6)              &      (7)          &  (8)    &(9)&  (10)  &  (11)       &  (12)  \\
\hline

X-01-00& 2000-03-29& 51632.97& $1.738^{\pm0.013}$ &$1.133^{\pm0.008}$  &76.83/81&$0.029^{\pm0.0005}$& $0.336^{\pm0.011}$& $379.89^{\pm15.21}$& $3.394^{\pm0.131}$& $6.998^{\pm0.099}$& 71.06/71\\ 
X-02-00& 2000-03-31& 51634.62& $1.710^{\pm0.013}$ &$1.197^{\pm0.007}$  &65.86/81&$0.018^{\pm0.0004}$& $0.346^{\pm0.014}$& $387.17^{\pm15.69}$& $3.579^{\pm0.142}$& $6.978^{\pm0.101}$& 59.62/71\\ 
X-05-00& 2000-04-04& 51638.84& $1.709^{\pm0.006}$ &$1.318^{\pm0.004}$  &78.84/81&$0.022^{\pm0.0004}$& $0.347^{\pm0.009}$& $386.13^{\pm13.87}$& $3.608^{\pm0.127}$& $6.617^{\pm0.107}$& 80.27/71\\ 
Y-01-00& 2000-04-08& 51642.57& $1.716^{\pm0.005}$ &$1.395^{\pm0.004}$  &97.18/81&$0.023^{\pm0.0008}$& $0.347^{\pm0.013}$& $385.09^{\pm15.18}$& $3.570^{\pm0.103}$& $6.964^{\pm0.121}$& 74.12/71\\ 
Z-01-01& 2000-04-13& 51647.62& $1.709^{\pm0.005}$ &$1.387^{\pm0.003}$  &123.3/81&$0.021^{\pm0.0003}$& $0.346^{\pm0.011}$& $383.67^{\pm14.65}$& $3.640^{\pm0.108}$& $7.062^{\pm0.108}$& 104.0/71\\ 
Z-02-00& 2000-04-15& 51649.33& $1.706^{\pm0.007}$ &$1.401^{\pm0.004}$  &83.92/81&$0.021^{\pm0.0005}$& $0.346^{\pm0.014}$& $382.94^{\pm13.49}$& $3.640^{\pm0.112}$& $7.027^{\pm0.100}$& 74.61/71\\ 
Y-02-00& 2000-04-18& 51652.54& $1.709^{\pm0.005}$ &$1.481^{\pm0.004}$  &105.0/81&$0.021^{\pm0.0005}$& $0.348^{\pm0.015}$& $390.46^{\pm13.54}$& $3.619^{\pm0.156}$& $7.150^{\pm0.099}$& 98.41/71\\ 
Z-03-00& 2000-04-21& 51655.18& $1.718^{\pm0.008}$ &$1.446^{\pm0.006}$  &95.51/81&$0.022^{\pm0.0009}$& $0.346^{\pm0.012}$& $394.87^{\pm14.66}$& $3.556^{\pm0.158}$& $7.291^{\pm0.107}$& 86.16/71\\ 
Z-03-01& 2000-04-24& 51658.86& $1.698^{\pm0.009}$ &$1.624^{\pm0.007}$  &59.70/81&$0.021^{\pm0.0004}$& $0.353^{\pm0.008}$& $392.46^{\pm13.44}$& $3.665^{\pm0.127}$& $7.392^{\pm0.116}$& 56.45/71\\ 
Z-03-02& 2000-04-27& 51661.09& $1.710^{\pm0.008}$ &$1.511^{\pm0.006}$  &83.68/81&$0.022^{\pm0.0006}$& $0.345^{\pm0.009}$& $390.04^{\pm17.36}$& $3.632^{\pm0.139}$& $7.119^{\pm0.134}$& 81.17/71\\ 
Y-03-00& 2000-04-29& 51663.20& $1.714^{\pm0.003}$ &$1.515^{\pm0.002}$  &144.7/81&$0.031^{\pm0.0005}$& $0.335^{\pm0.009}$& $402.95^{\pm15.72}$& $3.707^{\pm0.130}$& $7.273^{\pm0.111}$& 107.1/71\\ 
Z-04-02& 2000-05-01& 51665.48& $1.718^{\pm0.006}$ &$1.511^{\pm0.005}$  &118.9/81&$0.030^{\pm0.0005}$& $0.328^{\pm0.011}$& $383.29^{\pm13.88}$& $3.735^{\pm0.149}$& $6.552^{\pm0.121}$& 102.4/71\\ 
Z-04-01& 2000-05-04& 51668.22& $1.719^{\pm0.008}$ &$1.437^{\pm0.005}$  &88.88/81&$0.032^{\pm0.0007}$& $0.321^{\pm0.014}$& $404.50^{\pm12.12}$& $3.719^{\pm0.143}$& $6.595^{\pm0.101}$& 81.76/71\\ 
Z-05-01& 2000-05-06& 51670.94& $1.705^{\pm0.009}$ &$1.547^{\pm0.008}$  &92.80/81&$0.033^{\pm0.0007}$& $0.324^{\pm0.013}$& $427.81^{\pm14.11}$& $3.820^{\pm0.159}$& $7.405^{\pm0.117}$& 96.93/71\\ 
Z-05-02& 2000-05-08& 51672.46& $1.703^{\pm0.013}$ &$1.534^{\pm0.011}$  &74.82/81&$0.029^{\pm0.0008}$& $0.325^{\pm0.010}$& $418.73^{\pm13.89}$& $3.835^{\pm0.129}$& $7.232^{\pm0.100}$& 70.74/71\\ 
Z-05-03& 2000-05-10& 51674.52& $1.712^{\pm0.008}$ &$1.312^{\pm0.006}$  &93.47/81&$0.032^{\pm0.0006}$& $0.322^{\pm0.009}$& $408.21^{\pm13.44}$& $3.750^{\pm0.139}$& $6.582^{\pm0.114}$& 84.50/71\\ 
Z-05-04& 2000-05-11& 51675.66& $1.703^{\pm0.007}$ &$1.322^{\pm0.005}$  &86.03/81&$0.030^{\pm0.0003}$& $0.324^{\pm0.010}$& $428.67^{\pm12.36}$& $3.837^{\pm0.143}$& $6.866^{\pm0.116}$& 82.17/71\\ 
Z-06-00& 2000-05-15& 51679.37& $1.705^{\pm0.007}$ &$1.318^{\pm0.004}$  &103.7/81&$0.031^{\pm0.0009}$& $0.319^{\pm0.013}$& $425.65^{\pm12.72}$& $3.802^{\pm0.128}$& $6.949^{\pm0.108}$& 96.32/71\\ 
Z-07-00& 2000-05-19& 51683.50& $1.719^{\pm0.006}$ &$1.216^{\pm0.004}$  &66.68/81&$0.034^{\pm0.0008}$& $0.313^{\pm0.014}$& $427.81^{\pm11.12}$& $3.776^{\pm0.144}$& $7.134^{\pm0.099}$& 62.11/71\\ 
Z-07-01& 2000-05-23& 51687.01& $1.720^{\pm0.006}$ &$1.185^{\pm0.003}$  &79.64/81&$0.033^{\pm0.0008}$& $0.319^{\pm0.011}$& $405.95^{\pm14.09}$& $3.703^{\pm0.137}$& $6.699^{\pm0.093}$& 75.87/71\\ 
Z-08-00& 2000-05-27& 51691.35& $1.730^{\pm0.006}$ &$1.206^{\pm0.004}$  &84.41/81&$0.036^{\pm0.0007}$& $0.310^{\pm0.009}$& $425.61^{\pm14.99}$& $3.727^{\pm0.119}$& $7.241^{\pm0.097}$& 81.34/71\\ 
Z-10-00& 2000-06-11& 51706.09& $1.719^{\pm0.006}$ &$1.159^{\pm0.004}$  &108.2/81&$0.032^{\pm0.0006}$& $0.308^{\pm0.014}$& $403.65^{\pm13.26}$& $3.855^{\pm0.130}$& $6.528^{\pm0.151}$& 103.5/71\\ 
Z-10-01& 2000-06-15& 51710.83& $1.738^{\pm0.009}$ &$1.144^{\pm0.005}$  &52.01/81&$0.034^{\pm0.0008}$& $0.306^{\pm0.014}$& $380.36^{\pm15.57}$& $3.676^{\pm0.126}$& $6.414^{\pm0.137}$& 46.54/71\\ 
Y-05-01& 2000-06-24& 51719.92& $1.723^{\pm0.006}$ &$1.132^{\pm0.003}$  &74.39/81&$0.034^{\pm0.0007}$& $0.316^{\pm0.013}$& $413.30^{\pm13.21}$& $3.722^{\pm0.118}$& $6.865^{\pm0.117}$& 68.34/71\\ 
Z-12-00& 2000-06-28& 51723.18& $1.724^{\pm0.007}$ &$1.129^{\pm0.004}$  &99.79/81&$0.036^{\pm0.0006}$& $0.311^{\pm0.011}$& $427.40^{\pm12.44}$& $3.732^{\pm0.111}$& $6.504^{\pm0.123}$& 101.0/71\\ 
Z-13-01& 2000-07-05& 51730.81& $1.732^{\pm0.007}$ &$0.952^{\pm0.003}$  &82.64/81&$0.038^{\pm0.0008}$& $0.306^{\pm0.011}$& $425.71^{\pm13.76}$& $3.757^{\pm0.140}$& $7.002^{\pm0.096}$& 89.81/71\\ 
Z-13-02& 2000-07-06& 51731.81& $1.732^{\pm0.007}$ &$0.947^{\pm0.004}$  &113.4/81&$0.038^{\pm0.0006}$& $0.305^{\pm0.009}$& $424.34^{\pm12.72}$& $3.772^{\pm0.122}$& $7.179^{\pm0.115}$& 105.7/71\\ 
Y-04-01& 2000-07-08& 51733.86& $1.737^{\pm0.007}$ &$0.906^{\pm0.005}$  &62.17/81&$0.038^{\pm0.0009}$& $0.311^{\pm0.011}$& $400.23^{\pm12.04}$& $3.639^{\pm0.139}$& $6.957^{\pm0.108}$& 57.28/71\\ 
Z-14-01& 2000-07-13& 51738.72& $1.733^{\pm0.006}$ &$0.815^{\pm0.003}$  &102.5/81&$0.038^{\pm0.0007}$& $0.312^{\pm0.009}$& $428.81^{\pm13.21}$& $3.687^{\pm0.117}$& $7.497^{\pm0.109}$& 101.6/71\\
Z-15-00& 2000-07-22& 51747.81& $1.756^{\pm0.005}$ &$0.362^{\pm0.002}$  &41.12/46&$0.042^{\pm0.0006}$& $0.289^{\pm0.008}$& $379.93^{\pm11.02}$& $3.311^{\pm0.111}$& $7.175^{\pm0.105}$& 40.18/42\\
Z-14-02& 2000-07-25& 51750.86& $1.751^{\pm0.006}$ &$0.205^{\pm0.003}$  &46.52/46&$0.044^{\pm0.0008}$& $0.286^{\pm0.007}$& $379.82^{\pm10.71}$& $3.269^{\pm0.131}$& $7.267^{\pm0.107}$& 30.85/42\\
Z-16-00& 2000-07-30& 51755.08& $1.736^{\pm0.006}$ &$0.088^{\pm0.003}$  &41.89/46&$0.044^{\pm0.0008}$& $0.284^{\pm0.007}$& $379.62^{\pm12.22}$& $3.290^{\pm0.112}$& $7.301^{\pm0.110}$& 30.85/42\\

\hline
\end{tabular}
\noindent{
\leftline{X=50137-01, Y=50133-01, Z=50407-01 are prefixes of observation IDs.}
\leftline{$\Gamma$ represents the photon indices obtained from pure PL model fitting. PL flux indicates the flux from PL model in $10^{-9}~erg~cm^{-2}s^{-1}$. }
\leftline{$\dot{m}_d$, $\dot{m}_h$, $X_s$, $R$, $M_{BH}$ are the TCAF fitted parameters. The accretion rates ($\dot{m}_d$ and $\dot{m}_h$) are in Eddington rate.}
\leftline{$X_s$ is the shock location values in $r_s$ unit. $R$ is the compression ratio and $M_{BH}$ represents the values of mass obtained from the fit in $M_\odot$.}
\leftline{PL and TCAF model fitted $\chi^2_{red}$ values are mentioned as $\chi^2$/dof in Cols. 6 \& 12 respectively, where `dof' represents the degrees of freedom.}
\leftline{The superscripts are average error values of $\pm$ 90\% confidence extracted using `err' task in XSPEC.}
}
\end{table}

\clearpage
\begin{table}
\small
\addtolength{\tabcolsep}{-3.0pt}
\caption{X-ray Flux Contributions of Total, Accretion disk, and Jets}
\label{tab:table2}
\begin{tabular}{lcc|ccc|c|c}
\hline
Obs Id. & Date & MJD & $F_x$ & $F_{inf}$ & $F_{ouf}$ & \% of $F_{ouf}$& Norm  \\
(1)     &(2)   & (3) &  (4)  & (5)       &  (6)      &  (7) & (8)       \\
\hline
X-01-00& 2000-03-29& 51632.97& $1.132^{\pm0.006}$ &$0.330^{\pm0.003}$ &$0.802^{\pm0.009}$ &70.84& $14.75^{\pm0.201}$\\
X-02-00& 2000-03-31& 51634.62& $1.196^{\pm0.005}$ &$0.307^{\pm0.004}$ &$0.889^{\pm0.009}$ &74.33& $16.77^{\pm0.198}$\\
X-05-00& 2000-04-04& 51638.84& $1.316^{\pm0.005}$ &$0.356^{\pm0.003}$ &$0.960^{\pm0.008}$ &72.94& $15.85^{\pm0.179}$\\
Y-01-00& 2000-04-08& 51642.57& $1.394^{\pm0.005}$ &$0.364^{\pm0.004}$ &$1.030^{\pm0.009}$ &73.88& $16.43^{\pm0.201}$\\
Z-01-01& 2000-04-13& 51647.62& $1.387^{\pm0.006}$ &$0.395^{\pm0.005}$ &$0.992^{\pm0.011}$ &71.52& $15.13^{\pm0.211}$\\
Z-02-00& 2000-04-15& 51649.33& $1.400^{\pm0.006}$ &$0.396^{\pm0.004}$ &$1.004^{\pm0.010}$ &71.71& $15.17^{\pm0.189}$\\
Y-02-00& 2000-04-18& 51652.54& $1.481^{\pm0.005}$ &$0.372^{\pm0.003}$ &$1.109^{\pm0.008}$ &74.88& $17.08^{\pm0.190}$\\
Z-03-00& 2000-04-21& 51655.18& $1.447^{\pm0.006}$ &$0.342^{\pm0.005}$ &$1.105^{\pm0.011}$ &76.36& $18.16^{\pm0.179}$\\
Z-03-01& 2000-04-24& 51658.86& $1.621^{\pm0.005}$ &$0.422^{\pm0.005}$ &$1.199^{\pm0.010}$ &73.96& $16.50^{\pm0.168}$\\
Z-03-02& 2000-04-27& 51661.09& $1.510^{\pm0.005}$ &$0.387^{\pm0.003}$ &$1.123^{\pm0.008}$ &74.37& $16.77^{\pm0.200}$\\
Y-03-00& 2000-04-29& 51663.20& $1.516^{\pm0.007}$ &$0.456^{\pm0.004}$ &$1.060^{\pm0.011}$ &69.92& $14.28^{\pm0.192}$\\
Z-04-02& 2000-05-01& 51665.48& $1.511^{\pm0.006}$ &$0.399^{\pm0.003}$ &$1.112^{\pm0.009}$ &73.59& $16.26^{\pm0.184}$\\
Z-04-01& 2000-05-04& 51668.22& $1.436^{\pm0.007}$ &$0.364^{\pm0.004}$ &$1.072^{\pm0.011}$ &74.65& $16.96^{\pm0.185}$\\
Z-05-01& 2000-05-06& 51670.94& $1.547^{\pm0.007}$ &$0.399^{\pm0.005}$ &$1.148^{\pm0.012}$ &74.20& $16.65^{\pm0.167}$\\
Z-05-02& 2000-05-08& 51672.46& $1.534^{\pm0.006}$ &$0.396^{\pm0.004}$ &$1.138^{\pm0.010}$ &74.18& $16.64^{\pm0.179}$\\
Z-05-03& 2000-05-10& 51674.52& $1.310^{\pm0.005}$ &$0.365^{\pm0.003}$ &$0.945^{\pm0.008}$ &72.13& $15.45^{\pm0.181}$\\
Z-05-04& 2000-05-11& 51675.66& $1.322^{\pm0.005}$ &$0.342^{\pm0.004}$ &$0.980^{\pm0.009}$ &74.13& $16.60^{\pm0.201}$\\
Z-06-00& 2000-05-15& 51679.37& $1.314^{\pm0.006}$ &$0.338^{\pm0.003}$ &$0.976^{\pm0.009}$ &74.27& $16.68^{\pm0.178}$\\
Z-07-00& 2000-05-19& 51683.50& $1.215^{\pm0.005}$ &$0.329^{\pm0.004}$ &$0.886^{\pm0.009}$ &72.92& $15.87^{\pm0.184}$\\
Z-07-01& 2000-05-23& 51687.01& $1.184^{\pm0.007}$ &$0.357^{\pm0.005}$ &$0.827^{\pm0.012}$ &69.84& $14.26^{\pm0.182}$\\
Z-08-00& 2000-05-27& 51691.35& $1.205^{\pm0.006}$ &$0.324^{\pm0.003}$ &$0.881^{\pm0.009}$ &73.11& $15.98^{\pm0.190}$\\
Z-10-00& 2000-06-11& 51706.09& $1.159^{\pm0.007}$ &$0.357^{\pm0.003}$ &$0.802^{\pm0.010}$ &69.19& $13.96^{\pm0.183}$\\
Z-10-01& 2000-06-15& 51710.83& $1.141^{\pm0.007}$ &$0.354^{\pm0.004}$ &$0.787^{\pm0.011}$ &68.97& $14.85^{\pm0.188}$\\
Y-05-01& 2000-06-24& 51719.92& $1.132^{\pm0.006}$ &$0.345^{\pm0.004}$ &$0.787^{\pm0.010}$ &69.52& $14.08^{\pm0.192}$\\
Z-12-00& 2000-06-28& 51723.18& $1.127^{\pm0.005}$ &$0.289^{\pm0.003}$ &$0.838^{\pm0.008}$ &74.35& $16.83^{\pm0.193}$\\
Z-13-01& 2000-07-05& 51730.81& $0.951^{\pm0.005}$ &$0.317^{\pm0.004}$ &$0.634^{\pm0.009}$ &66.66& $12.90^{\pm0.187}$\\
Z-13-02& 2000-07-06& 51731.81& $0.946^{\pm0.006}$ &$0.333^{\pm0.004}$ &$0.613^{\pm0.010}$ &64.79& $12.19^{\pm0.186}$\\
Y-04-01& 2000-07-08& 51733.86& $0.905^{\pm0.005}$ &$0.345^{\pm0.005}$ &$0.560^{\pm0.010}$ &61.87& $11.26^{\pm0.190}$\\
Z-14-01& 2000-07-13& 51738.72& $0.814^{\pm0.005}$ &$0.329^{\pm0.005}$ &$0.485^{\pm0.010}$ &59.58& $10.63^{\pm0.191}$\\
Z-15-00& 2000-07-22& 51747.81& $0.361^{\pm0.005}$ &$0.211^{\pm0.004}$ &$0.150^{\pm0.009}$ &41.55& $7.334^{\pm0.141}$\\
Z-14-02& 2000-07-25& 51750.86& $0.205^{\pm0.006}$ &$0.201^{\pm0.003}$ &$0.004^{\pm0.009}$ &1.951& $4.383^{\pm0.130}$\\
Z-16-00& 2000-07-30& 51755.08& $0.205^{\pm0.005}$ &$0.201^{\pm0.003}$ &$0.004^{\pm0.008}$ &1.951& $4.369^{\pm0.135}$\\

\hline
\end{tabular}
\noindent{
\leftline{X=50137-01, Y=50133-01, Z=50407-01 are prefixes of observation IDs.}
\leftline{Total ($F_X$), accretion disk ($F_{inf}$), and Jet ($F_{ouf}$) X-ray fluxes are in units of 
$10^{-9}~ergs~cm^{-2}~s^{-1}$ and they are calculated in $3-25$~keV PCA band.}
\leftline{TCAF model fitted normalization ($N$) values with errors are shown in Col. 8.}
\leftline{Note: average values of 90\% confidence $\pm$ values obtained using `err' task in XSPEC, 
are plotted as superscripts of fitted parameter values.}
}
\end{table}
\end{document}